\documentclass[twocolumn]{aastex62}
\usepackage{float,graphicx,amsmath,multirow,mathtools}
\usepackage{color}
\usepackage[version=4]{mhchem}
\usepackage{comment}
\usepackage{amsmath}
\usepackage[caption=false]{subfig}
\usepackage{booktabs}
\usepackage{comment}
\usepackage{longtable}

\usepackage{bm}

\newcommand{\cmnegtwo}{cm$^{-2}$}

\begin{document}

\title{\ce{CH3}-Terminated Carbon Chains in the GOTHAM Survey of TMC-1: Evidence of Interstellar \ce{CH3C7N}}
\author{Mark A. Siebert}
\affiliation{Department of Astronomy, University of Virginia, Charlottesville, VA 22904, USA}
\author{Kin Long Kelvin Lee}
\affiliation{Department of Chemistry, Massachusetts Institute of Technology, Cambridge, MA 02139, USA}
\author{Anthony J. Remijan}
\affiliation{National Radio Astronomy Observatory, Charlottesville, VA 22903, USA}
\author{Andrew M. Burkhardt}
\affiliation{Center for Astrophysics $\mid$ Harvard~\&~Smithsonian, Cambridge, MA 02138, USA}
\affiliation{Department of Physics, Wellesley College, 106 Central Street, Wellesley, MA 02481, U.S.A.}
\author{The GOTHAM Collaboration}
\author{Ryan A. Loomis}
\affiliation{National Radio Astronomy Observatory, Charlottesville, VA 22903, USA}
\author{Michael C. McCarthy}
\affiliation{Center for Astrophysics $\mid$ Harvard~\&~Smithsonian, Cambridge, MA 02138, USA}
\author{Brett A. McGuire}
\affiliation{Department of Chemistry, Massachusetts Institute of Technology, Cambridge, MA 02139, USA}
\affiliation{National Radio Astronomy Observatory, Charlottesville, VA 22903, USA}
\affiliation{Center for Astrophysics $\mid$ Harvard~\&~Smithsonian, Cambridge, MA 02138, USA}

\correspondingauthor{Mark A. Siebert, Brett A. McGuire}
\email{mas5fb@virginia.edu, brettmc@mit.edu}

\submitjournal{ApJ}
\accepted{October 20, 2021}

\begin{abstract}
We report a systematic study of all known methyl carbon chains toward TMC-1 using the second data release of the GOTHAM survey, as well as a search for larger species. Using Markov-Chain Monte Carlo simulations and spectral line stacking of over 30 rotational transitions, we report statistically significant emission from methylcyanotriacetylene (\ce{CH3C7N}) at a confidence level of $4.6\sigma$, and use it to derive a column density of ${\sim}10^{11}$ \cmnegtwo\ .  We also searched for the related species, methyltetraacetylene (\ce{CH3C8H}), and place upper limits on the column density of this molecule. By carrying out the above statistical analyses for all other previously detected methyl-terminated carbon chains that have emission lines in our survey, we assess the abundances, excitation conditions, and formation chemistry of methylpolyynes (\ce{CH3C2_nH}) and methylcyanopolyynes (\ce{CH3C2_{n-1}N}) in TMC-1, and compare those with predictions from a chemical model. Based on our observed trends in column density and relative populations of the $A$ and $E$ nuclear spin isomers, we find that the methylpolyynes and methylcyanopolyynes families exhibit stark differences from one another, pointing to separate interstellar formation pathways, which is confirmed through gas-grain chemical modeling with \texttt{nautilus}.

\end{abstract}
\keywords{Astrochemistry, ISM: molecules }

\section{Introduction}
One of the most well-studied sites of cold, interstellar carbon chemistry is a prestellar core within the Taurus Molecular Cloud complex commonly referred to as Taurus Molecular Cloud 1 (TMC-1) \citep{1978A&A....67..139C}.  The chemistry occurring in TMC-1 produces a suite of exotic molecules (by terrestrial standards) including unsaturated cyanopolyyne (\ce{HC2_{n+1}N}; $n=1,2,3,...$) and acetylenic (\ce{C2_nH2}) linear carbon chains. Among the variations on these abundant molecules are the symmetric top methylpolyynes (MPs) and methylcyanopolyynes (MCPs), which have the form \ce{CH3C2_nH} and \ce{CH3C2_{n-1}N}, respectively. These species have been known in the interstellar medium (ISM) since the discovery of methyl cyanide (\ce{CH3CN}) \citep{solomon_detection_1971} and methyl acetylene (\ce{CH3CCH}) \citep{irvine_detection_1981}. Toward TMC-1, methylpolyynes and methylcyanopolyynes as large as \ce{CH3C6H} and \ce{CH3C5N} have been detected and characterized  \citep{matthews_detection_1983,irvine_detection_1981,broten_detection_1984,walmsley_ch3c4h_1984,macleod_ch3c4h_1984,loren_ch3c4h_1984,snyder_detection_2006,remijan_detection_2006}.

Currently, the chemical formation of MPs and MCPs in TMC-1 is unconstrained. While it is possible that simple carbon addition reactions occurring on the surfaces of dust grains may account for the growth of these species toward cold, dark clouds \citep{Turner_translucent_2000}, it is also plausible that ion-molecule reaction pathways in the gas phase can account for their formation \citep{quan_gasphase_2007}. Because molecules with C$_{3v}$ symmetry have unique excitation properties that do not allow for easy interconversion between $A$ and $E$ symmetry states, the relative populations of their $K$-ladders can offer additional clues to their formation conditions (see e.g.\ \cite{minh_ch3cn_1992} \cite{askne_ch3cch_1984}, \cite{Willacy1993_gasgrain_EA}, \citep{Mendoza2018_hc_AE}), but to date, analysis of these populations has only been done for the shortest MPs and MCPs. Furthermore, comparing the relative column densities of the similarly structured cyanopolyynes has proven to be a useful tool in testing chemical models and assessing the mechanisms through which carbon chain species grow in the ISM \citep{loomis_investigation_2020,cernicharo_irc_chains_2017,burkhardt_detection_2018}.  

Obtaining a more complete understanding of their chemistry and setting constraints on the abundances and excitation conditions of the various methyl-terminated carbon chains requires exploring the extent to which these types of species can grow (i.e. how long of a C-chain can form).  Such observations would enable the  direct comparison of measured abundances with both grain-surface and gas-phase reaction pathways, especially when combined with a fully self-consistent excitation analysis of MPs and MCPs in TMC-1. In addition, these studies will set the overall limit to the detectability of longer chain MPs and MCPs based on both an improved understanding of the chemical formation pathway and on a new computational methodology to detect the weak signals coming from the larger C-chain species. 

The ongoing GOTHAM survey (GBT Observations of TMC-1: Hunting for Aromatic Molecules) along with a deep Q-band survey being conducted on the Yebes 40\,m radio telescope \citep{cernicharo2020_HC3O+} have thus far proven to be very useful for rigorous studies of exotic molecules in TMC-1. With many new detections of various carbon-bearing molecules, their isomers, and cyclic/polycyclic species, these projects have effectively expanded our knowledge of the interstellar molecular carbon reservoir before star formation occurs \citep{xue_detection_2020,loomis_investigation_2020,mcguire_detection_2018,mccarthy_interstellar_2020,mcguire_early_2020,mcguire_discovery_2021,cabezas2021_cululene,cernicharo2021_C3HCCH}.  In this work, we present a detailed analysis of \ce{CH3}-terminated carbon chains toward TMC-1 as well as the first detection of methylcyanotriacetylene (\ce{CH3C7N}, \cite{Chen1981_MCPs}) using the second data release of GOTHAM. We also search for methyltetraacetylene (\ce{CH3C8H}, \cite{Travers1998_MPs}) in our data, but do not find significant emission and thus report upper limits on its abundance.

In Section \ref{obs}, we outline the specifics of our observations and reduction methods. In Section \ref{methods} we describe our spectroscopic calculations for the targeted molecules and statistical analysis procedures. In Section \ref{results} we present our results for all targeted molecules and discuss their relative column densities and excitation physics. Finally, in Section \ref{modeling} we discuss how our results fit in to the current understanding of chemistry in TMC-1, and compare them with predictions of our chemical model.

\section{Observations}
\label{obs}
Observations obtained in this study were collected as part of the GOTHAM Survey.  GOTHAM is a large project on the 100m Green Bank Telescope (GBT) currently carrying out dedicated spectral line observations of TMC-1 covering almost 30 GHz of radio bandwidth at high sensitivity and ultra-fine spectral resolution. This work uses the second data release (DR2) of GOTHAM, which includes over 600 hours of observations targeting the cyanopolyyne peak (CP) of TMC-1, centered at $\alpha_\mathrm{J2000}$ = 04\fh41\fm42.5\fs, $\delta_\mathrm{J2000}$ = +25\arcdeg41\arcmin26.8\arcsec. A full description of DR2 specifications and our reduction pipeline is described \citep{mcguire_early_2020}, but put briefly, the spectra in this data set cover the entirety of the X-, K-, and Ka-receiver bands with nearly continuous coverage from 8.0 to 11.6 GHz and 18.0 to 33.5 GHz (25.6 GHz of total bandwidth). All spectra have a uniform frequency resolution of 1.4\,kHz (0.05--0.01\,km/s in velocity) and an RMS noise of ${\sim}2$--20\,mK, with the RMS gradually increasing toward higher frequency because of the lower integration times. Data reduction involved removal of RFI and artifacts, baseline continuum fitting, and flux calibration using complementary VLA observations of the source J0530+1331. Uncertainty from to this flux calibration is estimated at $\sim$20\%, and is factored into our statistical analysis described below \citep{mcguire_early_2020}.

\section{Methods \& Analysis}
\label{methods}
\subsection{Molecular spectroscopy}
\label{section:molspec}

\begin{deluxetable}{ccccccr}
    \tablecaption{Spectroscopic properties of stacked \ce{CH3C7N} lines covered by the GBT X-band receiver }
    \tablewidth{\columnwidth}
    \tablehead{
    \multicolumn{3}{c}{Transitions} & \colhead{Symm.} &\colhead{Frequency}   &\colhead{$E_{up}$}   &\colhead{$S_{ij}\mu^2$}\\
    \colhead{$J'\ \rightarrow\ J''$}&\colhead{$K$}&\colhead{$F'\ \rightarrow\ F''$}&  &\colhead{(MHz)}&\colhead{(K)}& \colhead{(D$^2$)}}
     \tabletypesize{\scriptsize}
    \startdata
$11\ \rightarrow\ 10$ & $1$ & $10\ \rightarrow\ 9$& $E $ &8243.824   &9.92   &355.325  \\
   &   & $11\ \rightarrow\ 10$& $E $ &8243.824   &9.92   &389.482  \\
   &   & $12\ \rightarrow\ 11$& $E $ &8243.836   &9.92   &426.877  \\
   & $0$ & $10\ \rightarrow\ 9$& $A $ &8243.851   &2.37   &358.286  \\
   &   & $11\ \rightarrow\ 10$& $A $ &8243.859   &2.37   &392.727  \\
   &   & $12\ \rightarrow\ 11$& $A $ &8243.865   &2.37   &430.435  \\
 $12\ \rightarrow\ 11$ & $1$ & $11\ \rightarrow\ 10$& $E $ &8993.262   &10.35   &391.696  \\
   &   & $12\ \rightarrow\ 11$& $E $ &8993.263   &10.35   &426.021  \\
   &   & $13\ \rightarrow\ 12$& $E $ &8993.272   &10.35   &463.320  \\
   & $0$ & $11\ \rightarrow\ 10$& $A $ &8993.293   &2.81   &394.435  \\
   &   & $12\ \rightarrow\ 11$& $A $ &8993.299   &2.81   &429.000  \\
   &   & $13\ \rightarrow\ 12$& $A $ &8993.304   &2.81   &466.560  \\
 $13\ \rightarrow\ 12$ & $1$ & $12\ \rightarrow\ 11$& $E $ &9742.699   &10.82   &428.012  \\
   &   & $13\ \rightarrow\ 12$& $E $ &9742.700   &10.82   &462.478  \\
   &   & $14\ \rightarrow\ 13$& $E $ &9742.708   &10.82   &499.692  \\
   & $0$ & $12\ \rightarrow\ 11$& $A $ &9742.733   &3.27   &430.560  \\
   &   & $13\ \rightarrow\ 12$& $A $ &9742.739   &3.27   &465.231  \\
   &   & $14\ \rightarrow\ 13$& $A $ &9742.743   &3.27   &502.666  \\
 $14\ \rightarrow\ 13$ & $1$ & $13\ \rightarrow\ 12$& $E $ &10492.136   &11.33   &464.286  \\
   &   & $14\ \rightarrow\ 13$& $E $ &10492.137   &11.33   &498.870  \\
   &   & $15\ \rightarrow\ 14$& $E $ &10492.144   &11.33   &536.010  \\
   & $0$ & $13\ \rightarrow\ 12$& $A $ &10492.173   &3.78   &466.667  \\
   &   & $14\ \rightarrow\ 13$& $A $ &10492.178   &3.78   &501.429  \\
   &   & $15\ \rightarrow\ 14$& $A $ &10492.181   &3.78   &538.758  \\
 $15\ \rightarrow\ 14$ & $1$ & $14\ \rightarrow\ 13$& $E $ &11241.572   &11.86   &500.524  \\
   &   & $15\ \rightarrow\ 14$& $E $ &11241.573   &11.86   &535.211  \\
   &   & $16\ \rightarrow\ 15$& $E $ &11241.579   &11.86   &572.284  \\
   & $0$ & $14\ \rightarrow\ 13$& $A $ &11241.612   &4.32   &502.759  \\
   &   & $15\ \rightarrow\ 14$& $A $ &11241.616   &4.32   &537.600  \\
   &   & $16\ \rightarrow\ 15$& $A $ &11241.619   &4.32   &574.839  \\ \enddata   
   \tablecomments{Only $K=0$ and $K=1$ are included here since they contribute the brightest transitions; however we still considered states as high as $K=13$ in our MCMC model and stacking procedure. Similarly, lines ranging from $J'=27$ to $J'=44$ are covered by the GBT K- and Ka-bands and included in our analysis, but not shown here. For the full list of transitions used for \ce{CH3C7N} and all other methyl carbon chains treated in this work, please refer to the linked Dataverse repository,\dataset[10.7910/DVN/K9HRCK]{\doi{10.7910/DVN/K9HRCK}}.}
\label{tab:spec_dat}
\end{deluxetable}

All the MPs and MCPs are prolate symmetric top molecules with C$_{3v}$ symmetry, meaning their rotational energy states are defined by the total angular momentum state $J$ in addition to its projection along the unique axis of rotation, denoted by the quantum number $K$. Depending on the value of $J$ and $K$, symmetric top molecules will either exhibit $A$ or $E$ symmetry due to the nuclear spin on the methyl group; and since radiative and collisional transitions between these symmetry states are strictly limited, their relative populations can be far from the expected thermal distribution. Because of this, we treat transitions from non-degenerate ($A_1, A_2$ symmetry) and doubly degenerate ($E$) levels separately by running independent MCMC fits. As such, we can compare the relative abundances of carbon chains in these energy states, similar to the separate treatment of the $K=0$ and $K=1$ components employed by \cite{snyder_detection_2006} and \cite{remijan_detection_2006}.

For molecules with C$_{3v}$ symmetry, the total statistical weight of rotational energy levels is the product of the typical $J$-degeneracy $g_J=2J+1$, the $K$-level degeneracy ($g_k=1$ for $K=0$ and $g_k=2$ for $K\neq0$), and the nuclear spin degeneracy $g_I$ \citep{Gordy_Cook}. The reduced statistical weight contributed by nuclear spin statistics of three identical hydrogen nuclei in a methyl group (each with spin $I=1/2$) can be computed with the following:

\begin{equation}
    g_I = \frac{1}{3}\frac{\left(4I^2+4I+3\right)}{\left(2I+1\right)^2}\mathrm,\quad\mathrm{for}\quad K=0,3,6,9,...
\end{equation}
\begin{equation}
    g_I = \frac{1}{3}\frac{\left(4I^2+4I\right)}{\left(2I+1\right)^2}\mathrm,\quad\mathrm{for}\quad K\neq0,3,6,9,...
\end{equation}

After combining all degeneracies, the states with $A$ symmetry are weighted 2:1 relative to those with $E$ symmetry, except in the case of $K=0$ which is weighted equally because of the lower factor of $g_k$. The rotational constants for the largest MCPs and MPs were measured by \cite{Chen_1998_MCPspecs} and \cite{Travers_1998_MPspecs}. To generate the spectroscopic data used for this work from these measurements, we used PGopher \citep{western_pgopher:_2017,western_automatic_2019}, which is able to account for the necessary symmetry and statistical considerations. The input files and line lists for all six species are available on the Harvard DataVerse\dataset[10.7910/DVN/K9HRCK]{\doi{10.7910/DVN/K9HRCK}}.

As an example, Table \ref{tab:spec_dat} summarizes the spectroscopic properties of the rotational transitions of \ce{CH3C7N} in X-band that were a part of this investigation. The transition quantum numbers, calculated rest frequencies (MHz), upper state energy level ($K$), transition line strengths S($J, K$) are presented. In the case of the MCPs, the nuclear spin on the nitrogen nucleus contributes hyperfine splitting to each rotational energy level. 

Methyl acetylene (\ce{CH3CCH}) was not included in our analysis as it does not have any transitions covered by the GOTHAM survey. Similarly, although the fundamental transition of methyl cyanide (\ce{CH3CN}) is included in our spectra, a lack of additional lines prevents us from performing a full characterization of this molecule. For both \ce{CH3CN} and \ce{CH3CCH}, we refer to previous works studying these molecules in TMC-1 (e.g.\ \cite{askne_ch3cch_1984,minh_ch3cn_1992}), and instead base our analysis on the longer species.

\subsection{MCMC modeling}
\label{mcmc}
In order to derive physical characteristics for the \ce{CH3}- polyyne and cyanopolyyne species in TMC-1, we use the same Markov-Chain Monte Carlo (MCMC) model employed in previous publications from the GOTHAM collaboration \cite{mcguire_early_2020, xue_detection_2020, mccarthy_interstellar_2020}. This procedure is discussed at length in \cite{loomis_investigation_2020}, but we will summarize it here as well. In short, the MCMC model calculates probability distributions and covariances for parameters describing the physical and excitation conditions of a molecule in TMC-1.  Based on recent observations performed with the 45\,m telescope at Nobeyama Radio Observatory \citep{dobashi_spectral_2018,dobashi_discovery_2019}, as well as our data \citep{loomis_investigation_2020}, emission from molecules in TMC-1 toward the cyanopolyyne peak display at least four individual velocity components. In our current model, we make the assumption that the emission from each velocity component is cospatial, similar to the approach adopted in \citet{loomis_investigation_2020} for the cyanopolyynes. To summarize, the model describing a given methyl chain comprises a source size, four radial velocities, eight column densities and two excitation temperatures describing A/E components, and a linewidth parameter, giving a total of 16 parameters.

Line profile simulations were performed using \mbox{\textsc{molsim}} \citep{lee_molsim_2020}. The MCMC simulations used wrapper functions in \textsc{molsim} to \textsc{arviz} \citep{kumar_arviz_2019} and \textsc{emcee} \citep{foreman-mackey_emcee_2013}; the former for analyzing the results of sampling, and the latter implements an affine-invariant MCMC sampler. As prior parameters, we used the marginalized posteriors from the corresponding cyanopolyyne chain (e.g. \ce{HC3N} and \ce{CH3C3N}). The prior distributions are approximated as normally distributed (i.e. $p(\theta)\sim N(\mu_\theta, \sigma_\theta)$ for parameter $\theta$) with modifications to the variance $\sigma_\theta$ as to avoid overly constrictive/influential priors. Convergence of the MCMC was confirmed using standard diagnostics such as the \citet{gelman_single_1992} $\hat{R}$ statistic, and by visually inspecting the posterior traces. With each of the six species studied in this work, we compute separate MCMC model fits for the $A$ and $E$ symmetry states so their relative populations could be derived directly through comparison of the posterior distributions.

\subsection{Spectral line stacking and matched filter}
Following the formalism of \cite{loomis_detecting_2018,loomis_investigation_2020}, we perform a combined velocity stack and matched filter analysis in order to verify that the results of the MCMC model are consistent with the data, and determine the statistical significance of molecular detections. The first step involves a noise-weighted sum in velocity space of all expected transitions in our spectra, excluding any interloping transitions from known molecules in TMC-1; however, this was not necessary for any of the transitions in this analysis as there were no such interlopers. Next, the resulting composite spectrum is passed through a matched filter, using the best-fit parameters of the MCMC to create a model stacked spectrum. The impulse response of the matched filter, measured in signal-to-noise ratio $\sigma$, is a representation of how well the MCMC model reproduces GOTHAM spectra. Generally, we adopt a lower limit of $5\sigma$ as a threshold for confirmation of a molecule in TMC-1.

\section{Results \& Discussion}
\label{results}
\subsection{Column densities and physical characteristics}
\begin{figure*}[t!]
    \centering
    \includegraphics[width=\textwidth]{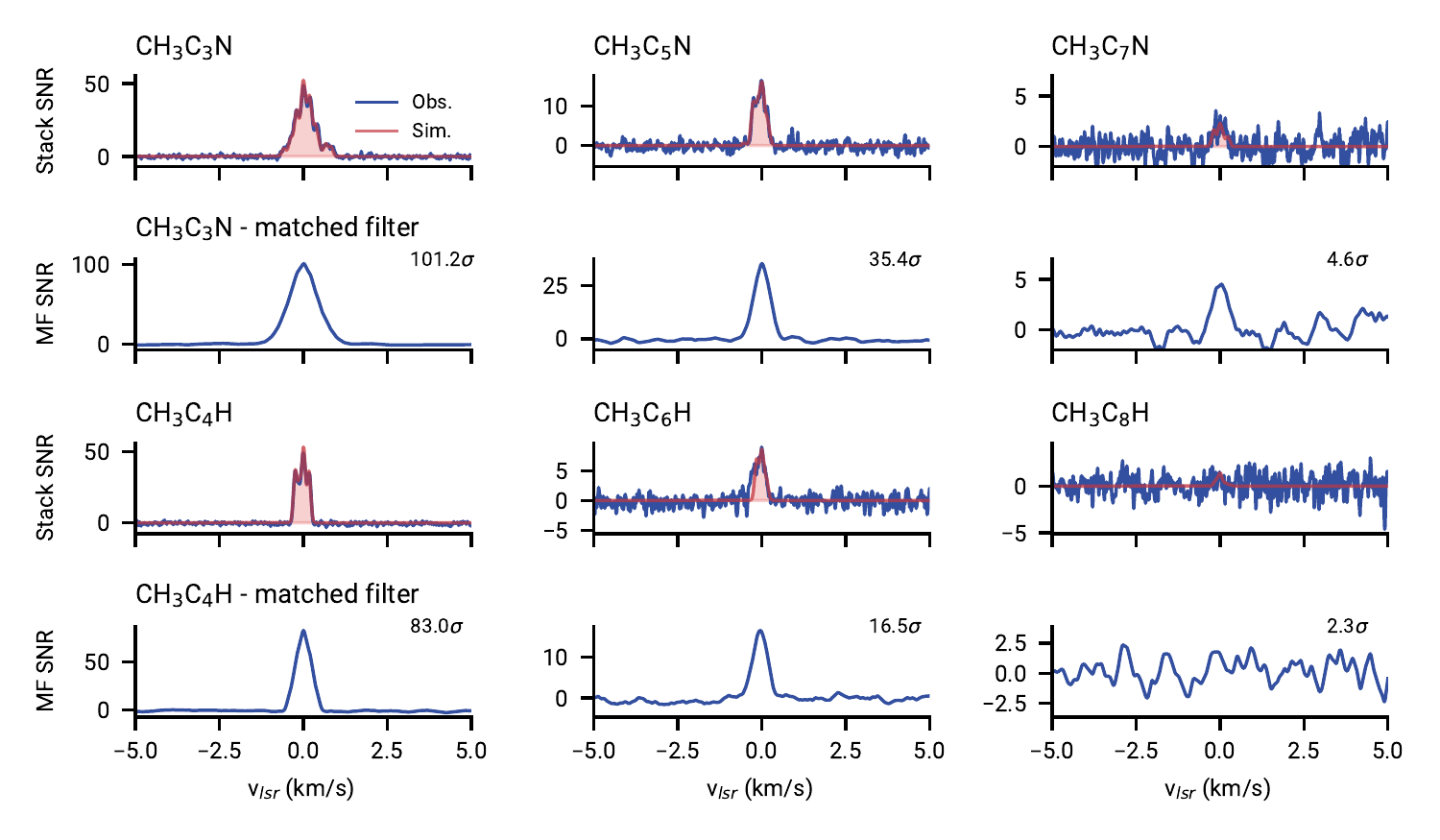}
    \caption{(\textit{First and third rows}) Velocity stacks of the observational (blue) and MCMC simulated (red) spectra. The cross-correlation of the two velocity stacks for each molecule corresponds to the matched filter (MF) spectrum (\textit{second and fourth rows}), which are shown in the second and fourth rows. Though the transitions belonging to $A$ and $E$ symmetry states were treated with separate MCMC fits, their results are combined here. The peak impulse response are annotated on the MF spectra.}
    \label{fig:stacks}
\end{figure*}

\begin{deluxetable*}{lcccccccc}[t!]
    \tablecaption{MCMC model results for all observed \ce{CH3}-terminated carbon chains. Uncertainties are listed as the 95\% confidence level.  }
    \tablehead{\colhead{Molecule} &\colhead{$N_T$} &\colhead{$N_A$} & \colhead{$N_E$} & \colhead{$T_{ex}$ ($A$ sym.)} & \colhead{$T_{ex}$ ($E$ sym.)} & \colhead{Source Size} & \colhead{$A/E$ Ratio}\\ - &  ($10^{11}$ cm$^{-2}$) &  ($10^{11}$ cm$^{-2}$) & ($10^{11}$ cm$^{-2}$) & (K) & (K) & $"$ & - }
    \startdata
    \ce{CH3C3N} & 8.66 (.46) & 5.00 (.40) & 3.66 (.24) & 3.36 (.40) & 5.70 (.26) & 481.0 (8.7) \tablenotemark{a} & 1.37 (.15)\\
    \ce{CH3C5N} & 2.86 (.30) & 2.01 (.29) & 0.85 (.07) & 3.48 (.34) & 3.97 (.48) & 128.7 (4.9) \tablenotemark{b} & 2.40 (.40)\\
    \ce{CH3C7N} & 0.86 (.19) & 0.69 (.18) & 0.17 (.06) & 3.82 (.25) & 3.90 (.44) &  53.9 (1.0) \tablenotemark{c} & 4.20 (1.8)\\
    \\
    \ce{CH3C4H} & 100.8 (5.7) & 60.0 (5.0) & 40.8 (2.8) & 4.31 (.84) & 5.50 (.37) & 481.0 (8.6) \tablenotemark{a} & 1.46 (.15)\\
    \ce{CH3C6H} & 10.4 (.72) & 6.10 (.60) & 4.30 (.40) & 7.19 (1.2) & 6.41 (1.1) & 128.3 (4.8) \tablenotemark{b} & 1.41 (.20)\\
    \ce{CH3C8H} & $<0.98$ & $<0.8$ & $<0.18$ & - & - & - & - \\
    \enddata
    \tablenotetext{a}{\ce{HC3N} source size adopted as prior. See \cite{loomis_investigation_2020} for more details.}
    \tablenotetext{b}{\ce{HC5N} source size adopted as prior.}
    \tablenotetext{c}{\ce{HC7N} source size adopted as prior.}
    \label{tab:mcmc}
\end{deluxetable*}

Figure~\ref{fig:stacks} shows the stacked data, the stacked MCMC model, as well as the matched filter response of all the methyl carbon chains considered in this work. We clearly detect \ce{CH3C3N}, \ce{CH3C5N}, \ce{CH3C4H}, and \ce{CH3C6H} at high significance, and the model is able to fit all components (both velocity and hyperfine) of these molecules. In the case of \ce{CH3C7N}, though no individual lines are present above the current noise level of the survey, the stacked emission in the top right of Figure~\ref{fig:stacks} exhibits noteworthy signal. Furthermore, its matched filter has a central peak at 4.6$\sigma$, indicating evidence that this molecule is present in our data, and that the MCMC model converged to a set of parameters that reproduce its emission. The matched filter response is weaker than previous molecules that were found using this same method \citep{loomis_investigation_2020,lee_CVA_2021,mcguire_discovery_2021} and falls just below our desired 5.0$\sigma$ threshold for a definitive discovery, but it is sufficient to consider a tentative detection and adopt the parameters of its fit in our analysis. For more information on this detection as it is indicated in the Bayesian fit to the spectra (prior to any stacking of emission lines), in Appendix \ref{cornerplots}, we present corner plots illustrating the full results of the MCMC model for both \ce{CH3C7N} and \ce{CH3C8H}, and discuss the posterior distributions and covariances in the parameter spaces for each. In contrast to \ce{CH3C7N}, \ce{CH3C8H} shows a much higher degree of uncertainty in its fit, in addition to no signal in the stack nor its matched filter response (bottom-right panels of Fig. \ref{fig:stacks}), so we therefore place upper limits on its column density.

A summary of parameters derived from the posterior distributions is provided in Table \ref{tab:mcmc}. Here, we note that the excitation temperatures of MPs (4.3 -- 7.2 K) are systematically higher than those of the MCPs (3.4 -- 3.9\,K), which is in agreement with previous studies of these species in TMC-1 \citep{broten_detection_1984,askne_ch3cch_1984,snyder_detection_2006}. However, one exception to this is seen for the $E$ state of \ce{CH3C3N}, which has $T_{ex}{\sim}5.7$ K. At all derived excitation temperatures, we find that less than 4\% of the total integrated line flux is contributed by transitions with $K>1$. Furthermore, the population of the $K=2$ state at these low temperatures is less than 1\% for all detected chains, which is well within the uncertainty of our measured column densities. In that sense, although states up to $K=13$ were considered, the column densities of the $A$ and $E$ components reported in Table \ref{tab:mcmc} can be treated as equivalent to the column densities of the $K=0$ and $K=1$ components, respectively.

\begin{figure}
    \centering
    \includegraphics[width=\linewidth]{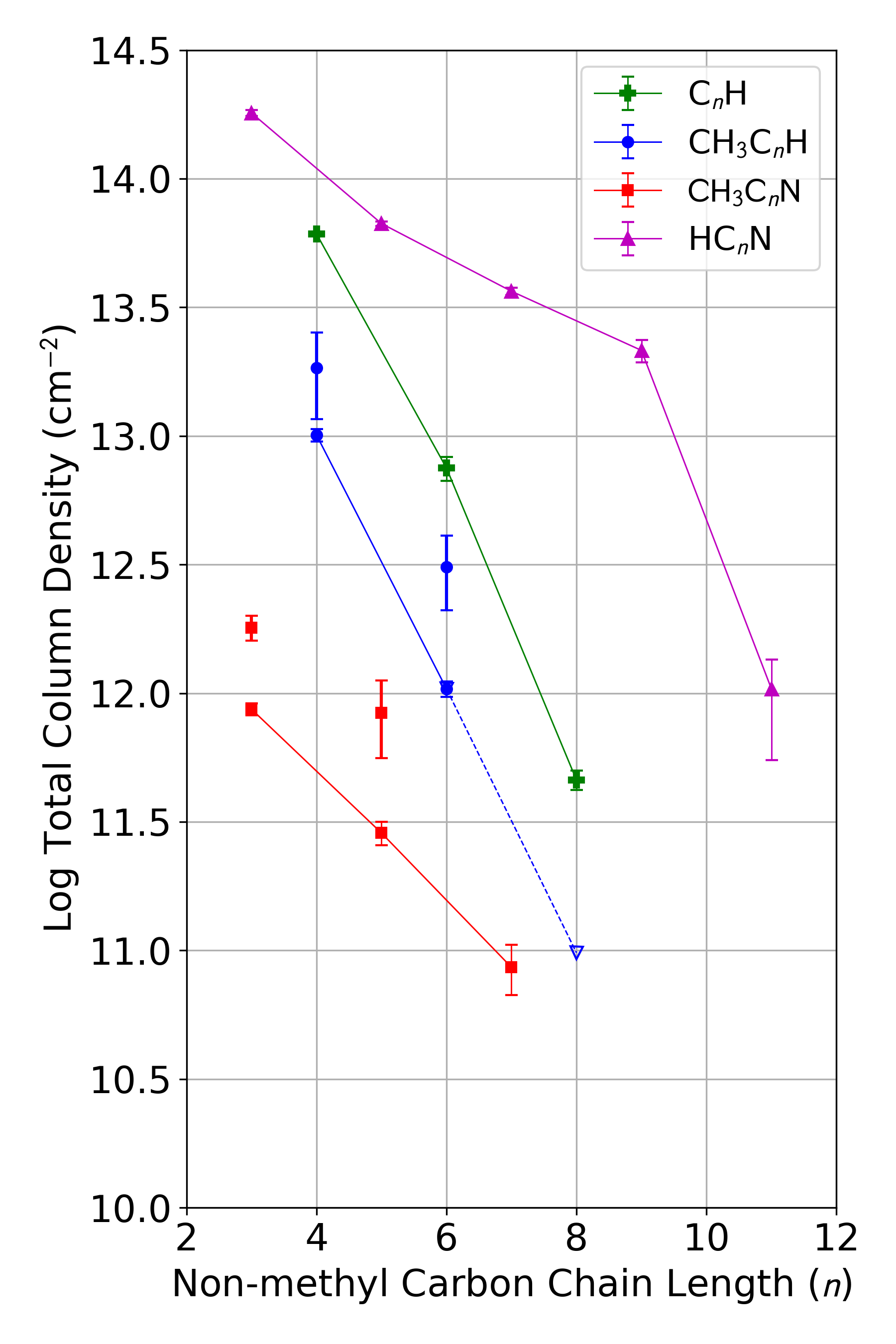}
    \caption{Derived total column densities of methyl-terminated carbon chains in TMC-1 as a function of chain length. Methylcyanopolyynes are shown in red while methylpolyynes are shown in blue. Contributions from the A and E symmetries of each species are summed, with their errors added in quadrature. \ce{CH3C8H} is denoted with a triangle as it is an upper limit. Results from \cite{remijan_detection_2006} are overplotted in their respective colors with no connecting line. Additionally, column  densities  of  the  linear  cyanopolyynes and hydrocarbons radicals published in \cite{loomis_investigation_2020} and \cite{Brunken_C8H-_detection} are shown in pink and green, respectively.}
    \label{fig:lengths}
\end{figure}

The total column densities (combining both $A$ and $E$ symmetries) decrease with increasing carbon chain length, much like the similar linear cyanopolyynes \citep{loomis_investigation_2020}. This is shown pictorially in Figure \ref{fig:lengths}, with a comparison to the column densities found by \cite{remijan_detection_2006}. There is small but notable disagreement (less than a factor of four) between the column densities we find and those derived in \cite{remijan_detection_2006}. We attribute this discrepancy to the intrinsic line strengths ($S_{ij}$) adopted from our numerical treatment of these molecules (see Section \ref{section:molspec}) as they are factors of 2--3 larger than the analytic values employed in \cite{remijan_detection_2006} and \cite{snyder_detection_2006}. Another important factor is that the column densities we derive take into account decreasing source sizes, whereas previous studies of TMC-1 assume no effects due to beam dilution. 

\ce{CH3C7N} has a total best-fit column density of $8.60 \pm 1.9 \times10^{10}$ \cmnegtwo, which matches well with the log-linear slope set by \ce{CH3C3N} and \ce{CH3C5N} (Fig.\ \ref{fig:lengths}). Furthermore, this slope is very similar to that observed for the cyanopolyynes, and would imply the next chain (\ce{CH3C9N}) to have a column density of about $3\times10^{10}$ \cmnegtwo. To detect stacked emission from this molecule at such an abundance, the RMS noise of the survey would need to be reduced by at least factor of five. This would require several hundred more hours of integration time, primarily at lower frequencies (e.g.\ the GBT X-band receiver) because the brightest transitions of this molecule are simulated to be present in the 8--10 GHz range.

In contrast to the cyanopolyynes and MCPs, the MPs decrease in abundance at a much faster rate than the other carbon chains in TMC-1 (about 1 dex for each subsequent species). Furthermore, the upper limit we place on \ce{CH3C8H} appears to break the trend seen in the shorter methylpolyynes. A linear extrapolation on the column densities of \ce{CH3C4H} and \ce{CH3C6H} (the blue points in Fig. \ref{fig:lengths}) would imply the next chain length has a column density ${\sim}2\times10^{11}$ \cmnegtwo, but in this case we would have detected emission from \ce{CH3C8H} at the sensitivity of our survey. Instead, it appears that the methylpolyynes experience a sharp drop-off in column density after \ce{CH3C6H}, which indicates that the production of these carbon chains is severely hindered for these longer species. 

A comparison to analogous linear species in TMC-1 is more difficult for the MPs, because the acetylene chains H$_2$C$_{2n}$ have no dipole moment, and thus no rotational transition lines with which we could apply a similar analysis. However, the hydrocarbon radicals (C$_{2n}$H) do exhibit transitions in the radio/sub-mm and have been characterized in TMC-1 up to a chain length of eight by \cite{Brunken_C8H-_detection} (green points in Fig. \ref{fig:lengths}). Here, it is apparent that those species also exhibit a non-linear decrease in column density after a carbon chain length of six. In other words, a similar ratio of $\frac{[\mathrm{\ce{CH3C8H}}]}{[\ce{CH3C6H}]}$ to what is observed for $\frac{[\mathrm{\ce{C8H}}]}{[\ce{C6H}]}$ would be in agreement with the upper limits we set in this work. This may suggest that similar chemical routes govern the abundances of both the MPs and the hydrocarbon radicals. However, since our upper limit on \ce{CH3C8H} is still very close to the log-linear extrapolated column density, it will be imperative to continue studying this molecule with future more sensitive data releases from GOTHAM survey to identify how steep this drop-off is, and how similar it is to the relative column density of \ce{C8H}.

\subsection{A/E symmetry state populations}
\begin{figure*}[t!]
    \centering
    \includegraphics[width=0.85\linewidth]{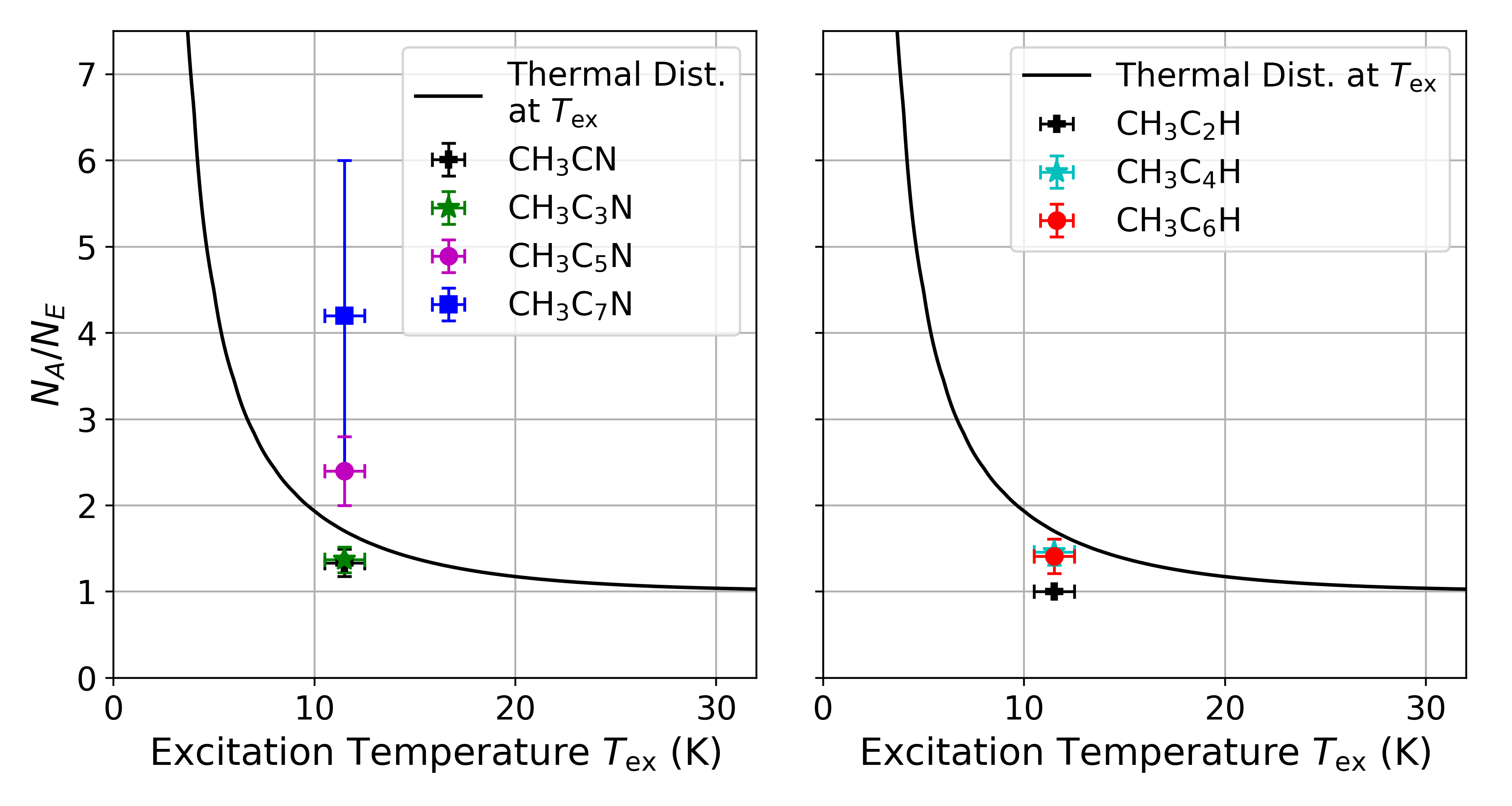}
    \caption{Column density ratios toward TMC-1 of $A$ and $E$ symmetry states for each of the chains studied in this work (colored points), as well as those derived for methyl cyanide and methyl acetylene by \cite{minh_ch3cn_1992} and \cite{askne_ch3cch_1984}, respectively (black points). Also plotted in black is the expected ratio of these two populations if the molecules are thermalized to a Boltzmann distribution at a range of kinetic temperatures. The vertical error on the points represents the 95\% confidence level from our MCMC modeling results, and the horizontal error bars denote the range of kinetic temperatures observed for gas phase molecules toward the CP region TMC-1 \citep{Feher_TMC_structure}. }
    \label{fig:AE_thermcomp}
\end{figure*}

\cite{askne_ch3cch_1984} found that in TMC-1, methyl acetylene (\ce{CH3CCH}) has about equal abundances of $A$ and $E$ symmetry states, which is not expected under LTE conditions and could indicate this molecule formed at a higher temperature. A similar result was found for \ce{CH3C4H} by \cite{walmsley_ch3c4h_1984}. In contrast, \cite{minh_ch3cn_1992} found that for \ce{CH3CN} in TMC-1, $N_E/N_A=0.76\pm0.09$, which suggests that these states are equilibrated to the kinetic temperature. With the sensitivity and bandwidth of GOTHAM DR2, we are able to revisit some of these population studies with a larger data set and expand them to the longer molecules \ce{CH3C6H}, \ce{CH3C3N}, \ce{CH3C5N}, and \ce{CH3C7N}. Because we performed separate MCMC analyses for the $A$ and $E$ states of each symmetric top carbon chain, their ratios can be computed directly and are shown in the last column of Table \ref{tab:mcmc}.

Figure \ref{fig:AE_thermcomp} shows a comparison of all $A/E$ ratios derived in this work plotted alongside the expected ratio in a thermalized distribution for a range of excitation temperatures. The gas phase kinetic temperature at the CP of TMC-1 has recently been placed at $11\pm1.0$\,K by \cite{Feher_TMC_structure}, so methyl chains with $A/E \approx 1.75$ can be considered as having equilibrated to this temperature.

For the MCPs (left plot in Fig.\ \ref{fig:AE_thermcomp}), we note that the $A/E$ ratio increases for longer chain lengths, starting at a superthermal value of 1.33 for \ce{CH3CN} (as measured by \cite{minh_ch3cn_1992}) and reaching $4.2\pm1.8$ for \ce{CH3C7N} which corresponds to an excitation temperature of $6.1\pm2$\,K, which is much lower than the kinetic temperature in TMC-1. In contrast, the MPs (right plot in Fig.\ \ref{fig:AE_thermcomp}) have nearly constant $A/E$ ratios between 1.4 and 1.5. While this is only slightly smaller than the equilibrium value of 1.75, it is indicative that all MPs studied here have slightly larger $E$ symmetry populations than what is expected given the physical conditions in TMC-1.

As noted in Section \ref{section:molspec}, both radiative and collisional interconversion between the $A$ and $E$ symmetry states are forbidden for both the MCPs and MPs (unless proton exchange collisions are non-negligible). \cite{Willacy1993_gasgrain_EA} investigated the possibility that $E\rightarrow A$ conversion of \ce{CH3CN} occurs through $E$-state species adsorbing to and subsequently desorbing from grain surfaces in TMC-1. They found that this is an efficient process so long as the desorption rate is high enough. This may suggest that an important distinction may be happening in TMC-1. Highly exothermic reactions in the gas phase (e.g. dissociative recombination) may drive the formation of the MPs, forcing them into superthermal $A/E$ distributions, where as the largest MCPs may preferentially form on (or stick to) the surfaces of dust grains. The non-thermal desorption of large complex molecules like the MCPs in TMC-1 has been a topic of numerous recent computation efforts \citep{Garrod2007_nontherm_desorp,Herbst2006_desorp,Hoang2019_desorp,Minissale2016_desorp,shingledecker:2021}, but it is still difficult to say exactly how efficient this process is for the species explored here. The investigation of the potential bifurcation of the formation pathways between the MPs and the MCPs using astrochemical models might shed further light on this intriguing situation.

\section{Chemical modeling of methyl chains}
\label{modeling}

To further explore the chemistry of methyl-terminated carbon chains in TMC-1, we utilized the adapted three-phase gas-grain chemical network model \texttt{nautilus} v1.1 code \citep{Ruaud:2016} discussed in previous analysis of GOTHAM data, which has previously been used to successfully study the formation of carbon-chain molecules \citep{xue_detection_2020,mcguire_early_2020,shingledecker:2021}. The physical conditions of the model are equal to what has been previously used with this network ($T_{\mathrm{gas}}=T_{\mathrm{grain}}=10$\,K, $n_{\text{H}_2}$=2$\times$10$^4$\,cm$^{-3}$, $A_V$=10, and  $\zeta_{\text{CR}}$=1.3$\times$10$^{-17}$\,s$^{-1}$; \citet{hincelin_oxygen_2011}) as are the elemental abundances \citep{loomis_investigation_2020}. Based originally off of the KIDA network, the network already contained the formation of MCPs up to $n=7$ and MPs up to $n=6$. Here, we expanded on this to include the formation of \ce{CH3C8H}, as well as additional reaction to better constrain the formation of the smaller members of this family.

To simulate the formation and destruction of \ce{CH3C8H}, we utilized the analogous pathways for the existing network of \ce{CH3C6H}, whose primary formation routes are from the dissociate recombination of \ce{C7H5+}. As such, we expanded out the formation of \ce{C9H5+} from ion-neutral reactions with semi-saturated carbon chains, whose rates were estimated using the Langevin formula \citep{woon_quantum_2009}. It should be noted that the rates for \ce{C7H5+} and \ce{CH3C6H} are calculated using the Modified Arrhenius formula, which may result in some potential discrepancies between this and \ce{CH3C8H}. As a source of destruction of \ce{C9H5+}, we adapted existing dissociate recombination of similar cations. The rates and branching ratios \ce{C9H5+} and \ce{C7H5+} were estimated and updated, respectively, based on the dissociate recombination of \ce{C5H5+} whose rates are taken from \citet{Herbst_Leung:1989} and branching ratios are discussed in detail on the KIDA website\footnote{\url{http://kida.astrophy.u-bordeaux.fr/datasheet/datasheet_2761_C5H5++e-_V1.pdf}}. In addition to dissociative recombination, C$_n$\ce{H5+} ions present in KIDA were also destroyed by reactions with anions. As such, the rates of the anion destruction of \ce{C9H5+} were based off analogous rates estimated for \ce{C7H5+} by \citet{Harada:2008}. Due to a lack of robust isomerization for large molecules within existing chemical networks, we only consider the  \ce{C7H5+} and \ce{C9H5+} isomers with KIDA entries (i.e. \ce{CH3C$_n$CH2+})

The analogous destruction of \ce{CH3C6H} was also adapted for \ce{CH3C8H}. This included 1) dissociation due to photons and cosmic rays 2) ion-neutral reactions with abundant ions (i.e.  \ce{H+}, \ce{H3+}, \ce{C+}, \ce{HCO+}, and \ce{He+}) which were estimated with Langevin formula and 3) reaction with elemental carbon to form \ce{C10H2} and \ce{H2}. To properly study \ce{C9H5+}, we also added in several neutral-neutral reactions to better account for the formation of \ce{C5H3}, a \ce{C9H5+}-precursor, which were obtained originally from \cite{Hebrard:2009}. Finally, we included an additional formation pathway for the MPs by reactions of the C$_{2n}$H family with methane based on work by \citet{quan_gasphase_2007} and formation of \ce{CH3C4H} from \ce{CH3CHCH2} by \citet{Berteloite:2010}.

For the MCPs, no additional reactions or rates were added from the KIDA network beyond the related adaptations from other recent molecular detections (e.g. the expanded semi-saturated carbon-chain networks for \ce{HC4NC} \citep{xue_detection_2020}, \ce{C6H5CN} \citep{Burkhardt2020_ubiquitous}, \ce{H2CCCHC3N} \citep{shingledecker:2021}, and indene \citep{Burkhardt:2021b}).

\begin{figure}
    \centering
    \includegraphics[width=\linewidth]{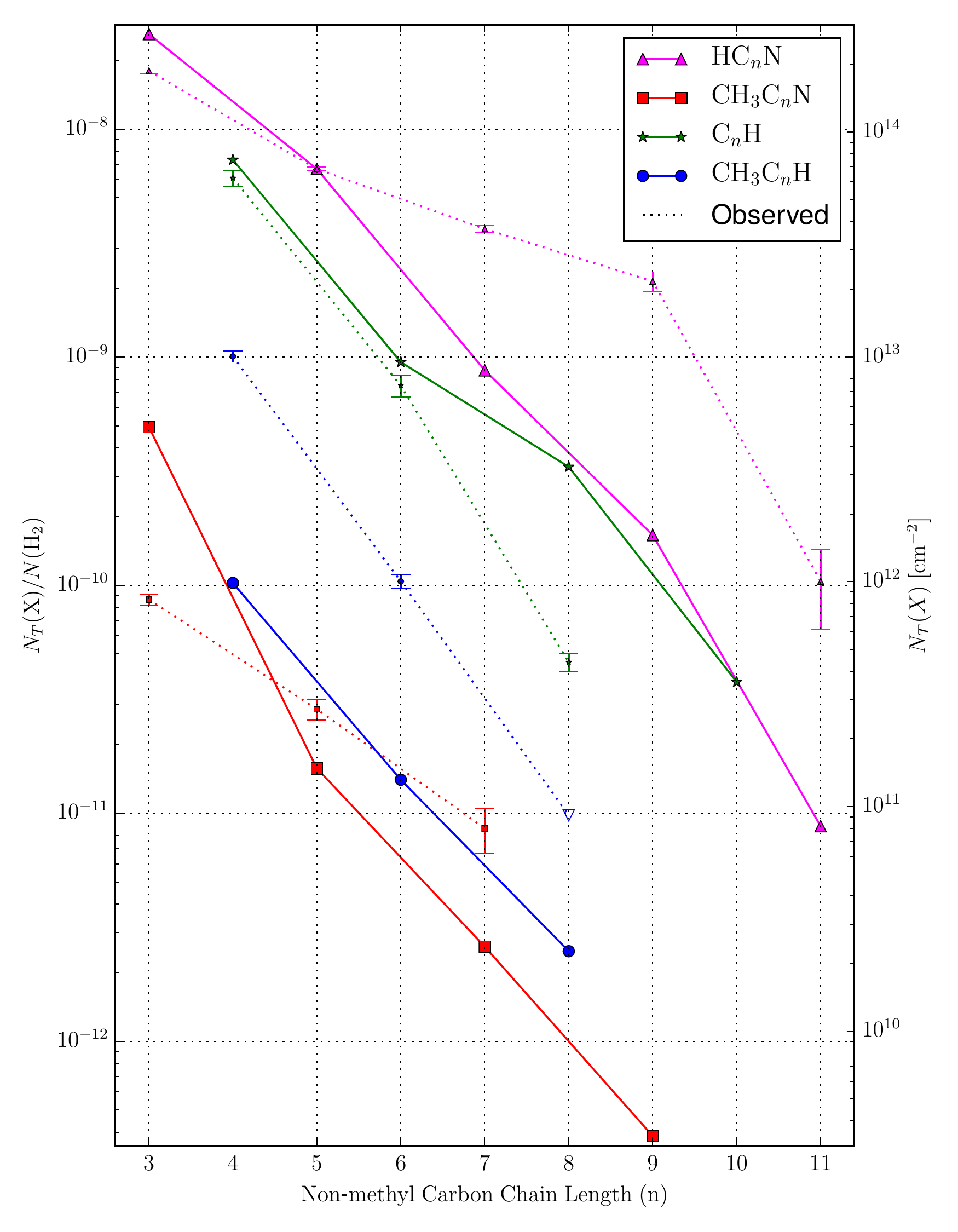}
    \caption{Simulated abundances and column densities of the four carbon chain families studied here at a model time of 2.5$\times$10$^5$ years, with the same colors and markers as Figure~\ref{fig:lengths}. Observed values are shown as dotted lines and the \ce{CH3C8H} upper limit is denoted as an unfilled triangle. }
    \label{fig:model}
\end{figure}

The results of this model can be seen in Figure~\ref{fig:model} at a time of 2.5$\times$10$^5$ yrs, assuming a TMC-1 hydrogen column density of $N_{\text{H}_2}$=10$^{22}$\,cm$^{-2}$. At this time in the model, the simulated abundances of all studied species are within an order of magnitude of their observed values. For the MCPs (solid red line in Fig.\ \ref{fig:model}), the abundance of \ce{CH3C3N} is higher than the observed value, while the remainder of the family are somewhat below their observed value. As such, the log-linear trend is less well constrained here. The relative trend line was also fairly independent of time, with a slight increase of the longer chains at later times, as seen in Fig.~\ref{fig:model_vs_t}. This is expected as longer chains typically have their peak abundance several 10$^5$ years after similar smaller chains, as seen by the cyanopolyynes \citep{loomis_investigation_2020}. The dominant formation pathways for the MCPs are the dissociative recombination of the C$_{n+1}$\ce{H4N+} ion family, which are in turn primarily produced by reactions between the cyanopolyynes and \ce{CH3+}, with minor contributions from C$_{n+1}$\ce{H3N+} and N + C$_{n+1}$\ce{H5+} after several 10$^5$ years. In particular, \ce{CH3C3N} has an additional production from \ce{H3C4N+} + \ce{H2} that may account for it deviation from the trendline of the rest of the MCP family. The MCPs are primarily destroyed by ion-neutral reactions with abundant ions (i.e.  \ce{H+}, \ce{H3+}, \ce{C+}, \ce{HCO+}, and \ce{He+}). The dependence on the cyanopolyynes (solid pink line in Fig.\ \ref{fig:model}) is consistent with the similar observed abundance trends seen in Figure~\ref{fig:lengths}. 

For the MPs (solid blue line in Fig.\ \ref{fig:model}), the relative abundance trends of both the simulated and observed values are in good agreement at this time. This slope of this trend is also very consistent with the simulated abundance trend of the C$_n$H family (solid green line in Fig.\ \ref{fig:model}). Within the general uncertainties of the source age of TMC-1 ($\sim$2-5$\times$10$^5$ years), the abundance of \ce{CH3C4H} was found to be consistently in worst agreement with the observed values but improving at later times by a $\sim$60\% abundance increase, while the longer chains increased up to nearly an order of magnitude (See Fig.~\ref{fig:model_vs_t}). As a result, the trend line was in less consistently in agreement with the observed trend at later times, while reproducing the observed abundances better. Due to the overall depletion of \ce{CH3C4H}, we have chosen this as a representative time as the simulated abundances are still within typical uncertainties for kinetic chemical models and we suspect the trend at this model time would remain consistent if a more efficient production is found either through new proposed pathways or improved reaction rate measurement/calculations. More generally, this strong dependence on the source age contrasts with other carbon chain families such as the cyanopolyynes and MCPs, whose observed trends are fairly consistent across the possible source ages of TMC-1. 

The dominant formation route for the MPs are from the dissociative recombination of the C$_{n+1}$\ce{H5+} family, with a significant contribution from C$_n$H reactions with methane after $\sim$10$^5$ years. The C$_{n+1}$\ce{H5+} family are produce by various ion neutral reactions, notably C$_{n}$\ce{H2+} with methane and C$_{n+1}$\ce{H2} with \ce{C2H4+} ($<10^5$ years). These dominant pathways provide strength for the observed relation between the MPs and C$_n$H families. There is also an increased importance after 10$^5$ years of \ce{C4H3+} + methane for \ce{CH3C4H} and \ce{C3H3} + \ce{C4H2} for \ce{CH3C6H}. MPs are mostly destroyed by ion-neutral reactions and reactions with carbon atoms. 

Overall, the predictions of these chemical models are within an order of magnitude of agreement with the observed values of TMC-1, indicating that carbon chain chemistry can be fairly well understood. In addition, the observed MP trend line can be reproduced well and is also analogous to what is observed for C$_n$H.  The estimated branching ratios and reaction rates still remain a major source of uncertainties in the models, which provide a strong motivation for further laboratory and theoretical studies. As discussed in \citet{Burkhardt:2021b}, detailed study of increasing larger molecules (e.g. molecules with 7+ heavy atoms and/or aromatic rings) will likely require a rigorous study of the formation of semi-saturated carbon chains that have historically been performed piecewise. The recent expanded inventory of known carbon chains in TMC-1 will provide key constraints in this study.

\section{Conclusions}
\label{summary}
We performed a rigorous, self-consistent study of methylpolyynes and methylcyanopolyynes in TMC-1 using the second data release of GOTHAM. Through MCMC modeling and matched filter analysis of stacked of emission lines, we derived column densities, excitation temperatures, and $A/E$ symmetry ratios for all previously found species, and discovered evidence for a new interstellar symmetric top, methylcyanotriacetylene (\ce{CH3C7N}), at a confidence level of 4.6$\sigma$. We also searched for methyltetraacetylene \ce{CH3C8H}) in our spectra and place upper limits on its column density in TMC-1.

In our analysis, we found two important divisions between the different families of molecules explored in this work:
\begin{enumerate}
    \item The column densities of methylcyanopolyynes decrease in a log-linear manner with increasing carbon chain length, and the slope of this trend is similar to the cyanopolyynes. In contrast, the methylpolyynes experience a drop-off in column density for species larger than \ce{CH3C6H} that is not consistent with the trend set by the smaller species.
    \item The $A/E$ ratios of methylcyanopolyynes increase with carbon chain length, and are subthermal for the larger species. The detected methylpolyynes have systematically smaller $A/E$ ratios which are not equilibrated to the kinetic temperature in TMC-1.
\end{enumerate}

Given the structural similarity of the two classes of molecules, these dichotomies are striking and point to separate formation and carbon chain growth mechanisms in TMC-1. Whether these differences exist between the analogous cyanopolyynes (HC$_{2n+1}$N) and the pure hydrocarbon polyynes (H$_2$C$_{2n}$) is unclear, as the latter have no dipole moment. To understand this, future infrared observations of vibrational bands from these molecules toward TMC-1 would be instrumental in constraining their abundances and understanding their formation chemistry.

Finally, utilizing a gas-grain chemical network, we modeled the formation and destruction of \ce{CH3}-terminated carbon chains in TMC-1. The model is able to reproduce our observed column densities to within an order of magnitude, and we see very similar trends in the exponentially decreasing abundances with carbon chain length. However, we are not able to predict the observed drop-off in column density for \ce{CH3C8H} or \ce{C8H} using this model.

The results of this analysis offer important insights into the formation of carbon chain molecules in cold pre-stellar conditions, but it is important to note that many other families of unsaturated carbon chain species are known to form in TMC-1 (e.g.\ C$_{n}$O, C$_{n}$S, HC$_{n}$O). Similar in-depth analyses of these groups of molecules would be instrumental in improving our understanding the extent and efficiency of carbon chemistry in TMC-1.

We thank the reviewers for their thoughtful notes on this manuscript. The National Radio Astronomy Observatory is a facility of the National Science Foundation operated under cooperative agreement by Associated Universities, Inc. The Green Bank Observatory is a facility of the National Science Foundation operated under cooperative agreement by Associated Universities, Inc. M.\,A.\,S.\,acknowledges support from the Virginia Space Grant Consortium for this work.

\bibliography{references}
\bibliographystyle{aasjournal}

\appendix

\renewcommand{\thefigure}{A\arabic{figure}}
\renewcommand{\thetable}{A\arabic{table}}
\renewcommand{\theequation}{A\arabic{equation}}
\setcounter{figure}{0}
\setcounter{table}{0}
\setcounter{equation}{0}

\section{MCMC posterior distributions}
\label{cornerplots}
\begin{figure}[h]
    \centering
    \includegraphics[width=\textwidth]{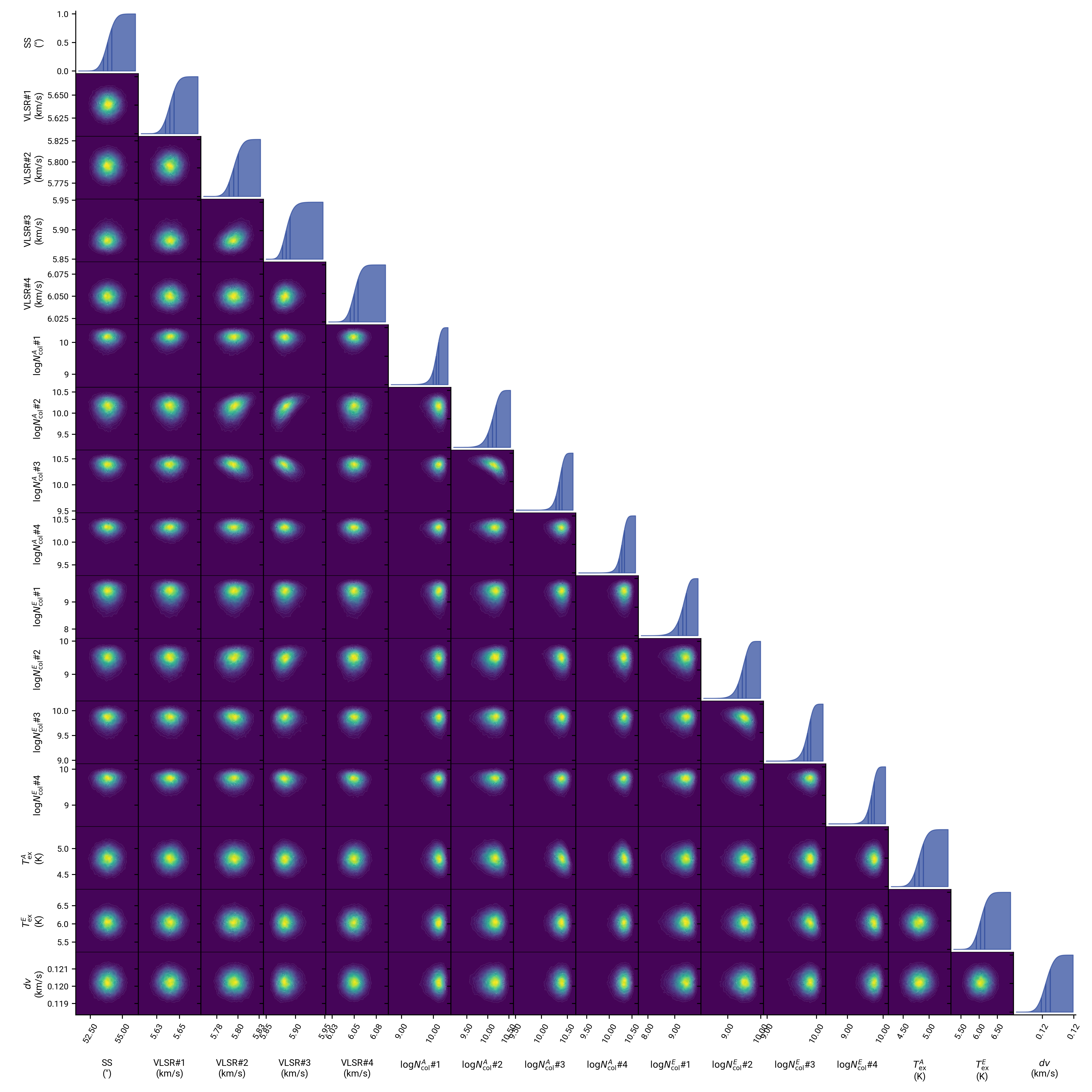}
    \caption{Corner plot for \ce{CH3C7N}. The diagonal traces correspond to the marginalized, cumulative posterior distributions for each parameter, and off-diagonal heatmaps represent parameter covariances. Vertical lines in the cumulative distribution plots represent the first, second, and third quantiles.}
    \label{fig:cornerplot_ch3c7n}
\end{figure}

\begin{figure}[h]
    \centering
    \includegraphics[width=\textwidth]{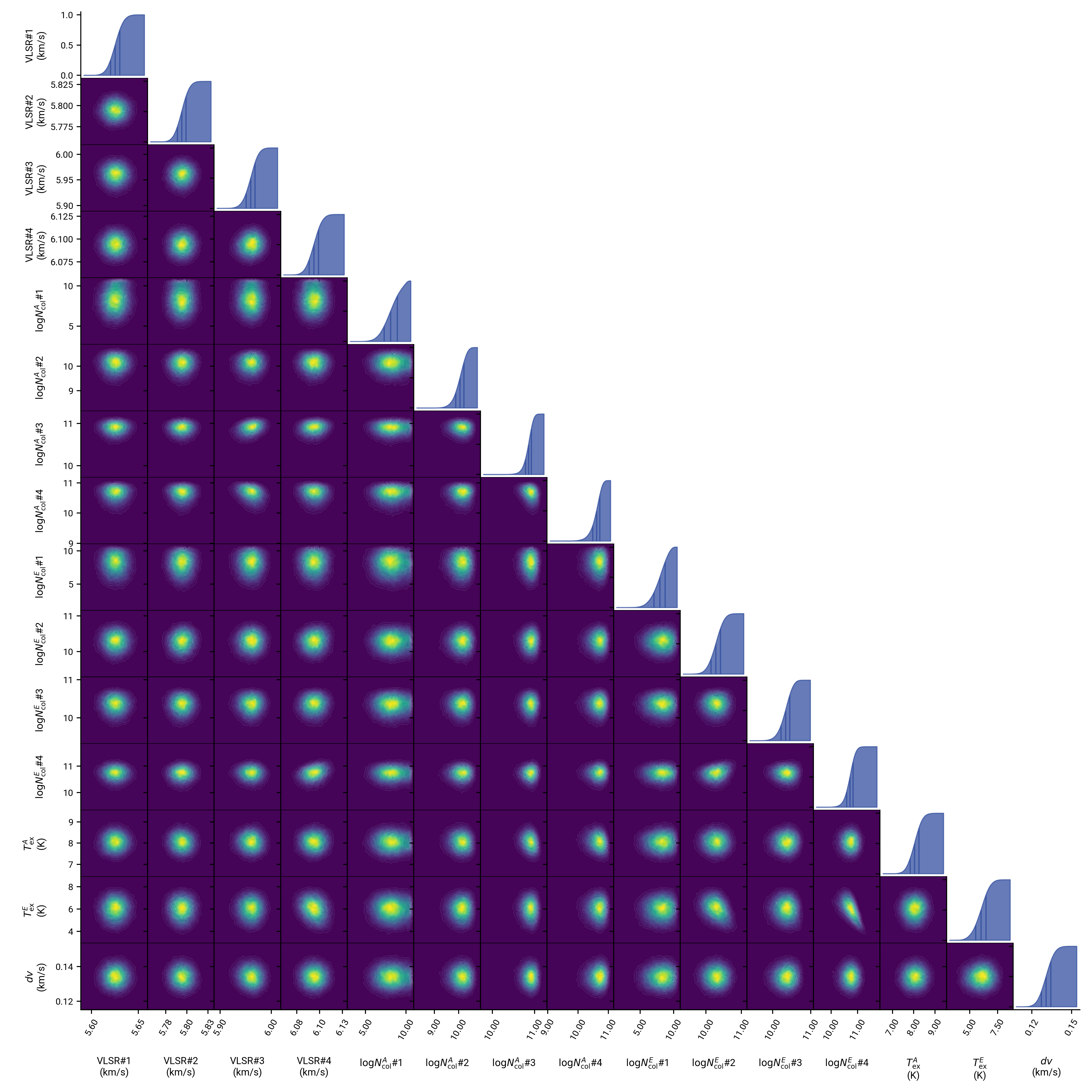}
    \caption{Corner plot for \ce{CH3C8H}. The diagonal traces correspond to the marginalized, cumulative posterior distributions for each parameter, and off-diagonal heatmaps represent parameter covariances. Vertical lines in the cumulative distribution plots represent the first, second, and third quantiles.}
    \label{fig:cornerplot_ch3c8h}
\end{figure}

Figure \ref{fig:cornerplot_ch3c7n} and \ref{fig:cornerplot_ch3c8h} shows the MCMC posterior distributions for \ce{CH3C7N} and \ce{CH3C8H} respectively. Based on the off-diagonal heatmaps, we see that the majority of modeling parameters do not demonstrate significant covariance; those that do typically pertain to radial velocities and column densities of each velocity component, often those adjacent to one another in velocity and within the same symmetry group (i.e. $E$ state column densities in neighboring velocity components). We do not expect significant covariance between the $A/E$ states for both species, as they are by in large spectroscopically separate.

Regarding the treatment of source sizes, we observe quite different posteriors in comparison to the earlier work done by \citet{loomis_investigation_2020}, which highlighted significant covariance between source sizes and column densities. We do not observe this for \ce{CH3C7N} most likely due to an overly constrained prior placed on the source size, and we likely underestimate its uncertainty and covariance. In the case of \ce{CH3C8H}, we fixed the mean value of \ce{HC7N} \cite{loomis_investigation_2020}, as we were interested in estimating upper limits to the column densities, and thus not shown in Figure \ref{fig:cornerplot_ch3c8h}.

Comparison of the posterior distributions for the column densities between \ce{CH3C7N} and \ce{CH3C8H} provides a margin for detection and non-detection cases. In the former, where the matched filtering shown in Figure \ref{fig:stacks} indicates a $4.6\sigma$ likelihood for \ce{CH3C7N}, corroborating with smaller uncertainties in the column densities in comparison to \ce{CH3C8H}, particularly for the first velocity component. We interpret this as a smaller range of column density values provide evidential support for the tentative detection of \ce{CH3C7N}: even with the absence of individual lines, the likelihood-based sampling is able to place relatively large constraints on the possible values.

\section{Chemical Model Time Dependence}
\begin{figure}[h]
    \centering
    \includegraphics[width=0.45\textwidth]{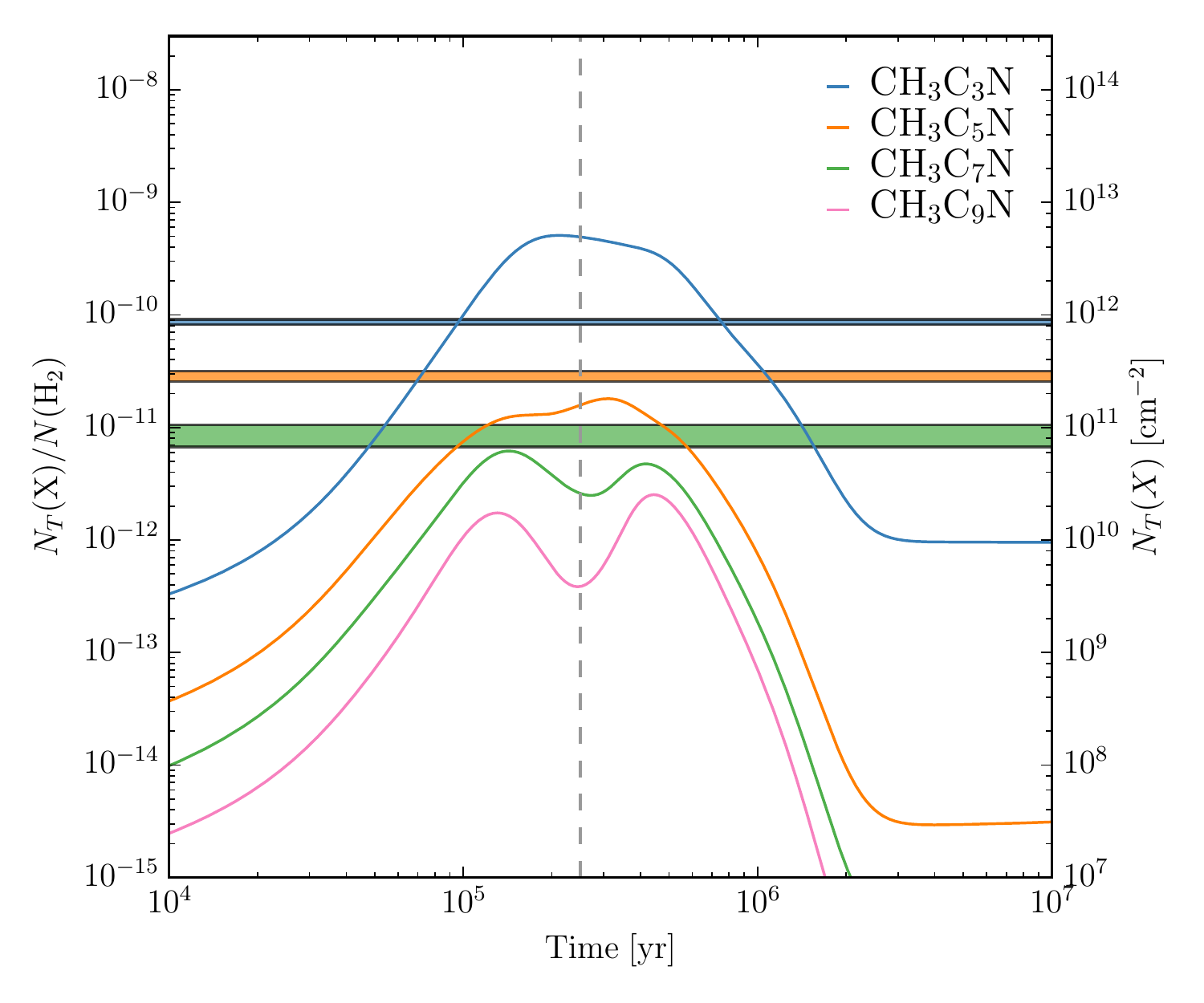}
    \includegraphics[width=0.45\textwidth]{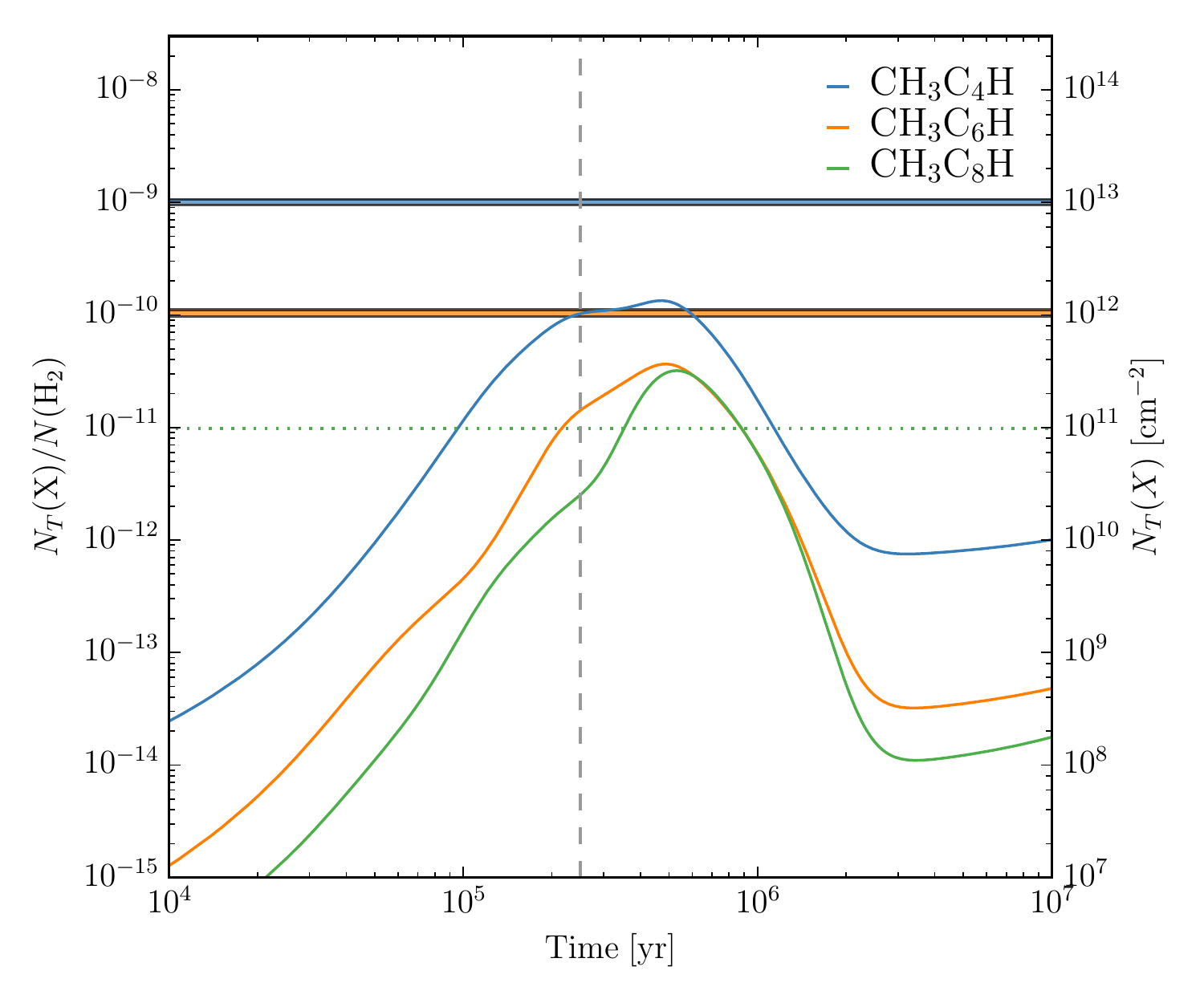}
    \caption{Simulated gas-phase abundance and column densities of the \ce{CH3C}$_n$\ce{N} (\textit{left}) and \ce{CH3C}$_n$\ce{H} (\textit{right}) families from \texttt{nautilus} chemical models in comparison to the observed values with uncertainties as a horizontal bars. The upper limit of the \ce{CH3C8H} is shown as green dotted line. The time used in Figure~\ref{fig:model} is shown as a vertical dashed gray line.}
    \label{fig:model_vs_t}
\end{figure}

Figure~\ref{fig:model_vs_t} shows the time-dependence of the simulated abundances within the \texttt{nautilus} chemical models in relation to the observed values. As it can be seen, the relative trend lines of these species can be quite time dependent, the time of peak abundance is strongly dependent on carbon chain length.

\end{document}